\documentclass[doublecol]{epl2}
\usepackage{amssymb,amsfonts,amsmath,graphics}
\usepackage{verbatim}
\newcommand{\be}{\begin{equation}}
\newcommand{\ee}{\end{equation}}
\newcommand{\tm}{\tau_{\mathrm{min}}}
\newcommand{\tM}{\tau_{\mathrm{max}}}

\title{Suppression of epidemic outbreaks with heavy-tailed contact dynamics}
\author{Byungjoon Min \and K.-I.~Goh\thanks{\email{kgoh@korea.ac.kr}} \and I.-M.~Kim} 
\shortauthor{B. Min \etal}
\institute{Department of Physics, Korea University, Seoul 136-713, Korea}

\abstract{
We study the epidemic spreading process following contact dynamics with heavy-tailed waiting time distributions.
We show both analytically and numerically that the temporal heterogeneity 
of contact dynamics can significantly suppress the disease's transmissibility, 
hence the size of epidemic outbreak, obstructing the spreading process. 
Furthermore, when the temporal heterogeneity is strong enough, one obtains 
the vanishing transmissibility for any finite recovery time and regardless of underlying structure of contacts, the condition of which was derived.
}

\begin{document}
\maketitle
\section{Introduction}
Throughout history, epidemics have had major influence not only on an individual's health but also on global human history \cite{epidemic},
proving to be of great intellectual and practical interest \cite{anderson}.
Over the last decade, classical epidemic theory had undergone a paradigm shift
in response to the recognition of nontrivial structure of social and technological networks over which the epidemics takes place \cite{vespig,newman,miller,kenah,castellano,gomez,goltsev,hklee}.
In particular, many real-world networks exhibit heterogeneous connectivity
structure, characterized by a heavy-tailed degree distribution $G(k)$ that often takes an asymptotic power-law form, $G(k)\sim k^{-\gamma}$, where the degree $k$ is the number of contacts (links) an agent (node) has \cite{network}. 
Such network heterogeneity has proven to have a major impact on epidemic dynamics \cite{vespig,newman}.
One of the most striking results in this regard is the vanishing epidemic threshold 
in scale-free (SF) networks with a power-law degree distribution 
with the degree exponent $\gamma\le3$ \cite{vespig,newman}.
This means that in such highly heterogeneous populations an 
epidemic can spread over the network even at an arbitrarily 
small infection rate, or equivalently, for infectious 
diseases with arbitrarily short lifetime (recovery time),
implying severe vulnerability of such systems to epidemic outbreaks.

More recently, the temporal heterogeneity of individual's activity has also been shown to influence dynamical processes on a network \cite{temporal}.
Various kinds of human activities including sexual activities \cite{browsing,rocha1,sex} exhibit strongly heterogeneous contact dynamics which is dominated by a few bursts of activities with extended periods of quiescence in between.
Such temporal heterogeneity \cite{temporal} can be dictated by the waiting time distribution $P(\tau)$,
where the waiting time $\tau$ is the time interval between two consecutive actions, and 
presents a new layer of complexity in social dynamics, 
parallel to the structural heterogeneity or network heterogeneity 
dictated by the complex network structure of contacts~\cite{network}.
Indeed, it has been shown that such temporal heterogeneity can
significantly affect spreading processes in networks~\cite{alexei,moro,karsai,min,lara,rocha1,mieghem,rocha2,lambiotte}.
In particular, it is found to be that non-exponential infection time distribution can significantly 
alter the epidemic threshold in susceptible-infected-susceptible model on networks \cite{mieghem}.
Furthermore, Rocha and collaborators investigated the effect of heavy-tailed activation dynamics in evolving networks on the epidemic spreading \cite{rocha2}.
In this paper, we investigate further the effect of temporal heterogeneity 
of contact dynamics on the large-scale properties of epidemics, focusing on the epidemic threshold.

The main result of this paper is to show analytically that the temporal heterogeneity 
can significantly impede the epidemic outbreak, in stark contrast 
with the network heterogeneity that facilitates it~\cite{vespig}.
The epidemic threshold can become arbitrarily large as the temporal heterogeneity diverges.
We demonstrate this analytically by applying renewal theory \cite{feller,vazquez}
to a prototypical epidemic model, the susceptible-infected-recovered (SIR) model. 
We derive expressions for the transmissibility and thus epidemic threshold
for heavy-tailed $P(\tau)$, to show that
the epidemic threshold increases with heterogeneity of contact dynamics without bound.
The analytical predictions are well supported by extensive
numerical simulations on random and scale-free networks.
We conclude the paper by discussing the role of finite cutoffs in waiting times.

\section{SIR model with arbitrary contact dynamics}
The SIR-type model we consider in this paper is formulated as follows.
A population of $N$ nodes is modeled as a network which is fixed in time (quenched).
Each node in the network is in one of three states, susceptible, infected, or recovered. 
Disease is transmitted upon a contact through the link between an infected node and its susceptible neighbor.
In classical approach \cite{anderson}, the contact dynamics is assumed to be a Poisson process. 
Here we relax such Poisson assumption, and consider that the contact dynamics through each link follows
independent renewal process with a general inter-event time (waiting time) distribution $P(\tau)$. 
Along the way, the infected node can recover autonomously, after when 
it does not participate in epidemic dynamics. 
Classically the recovery dynamics is also assumed to be Poissonian. 
In this work we consider a fixed recovery time $\lambda$, to focus on the effect of
heavy-tailed contact dynamics. 

Key quantities for disease spreading dynamics are the so-called 
transmissibility $T$ and the secondary reproductive number $R$ \cite{anderson,newman}. 
$T$ is the probability that an infected individual would transmit disease 
to a susceptible neighbor before it recovers, and 
$R$ is the expected number of secondary infections per each infected node. 
Given $P(\tau)$, $\lambda$ controls the ``infectiousness'' of the disease (and thus $T$).
If $\lambda$ is large (small), there is more (less) chance for secondary infections. 
Finally, the average fraction of recovered nodes $\rho_{\infty}$ 
at $t\to\infty$ could measure the expected ``size'' of epidemic outbreak. 
One defines the epidemic threshold $\lambda_c$ for the epidemic outbreak 
to be the infimum of $\lambda$ such that $\rho_{\infty}>0$. 

\begin{figure}[t]
\centering\includegraphics[width=.95\linewidth]{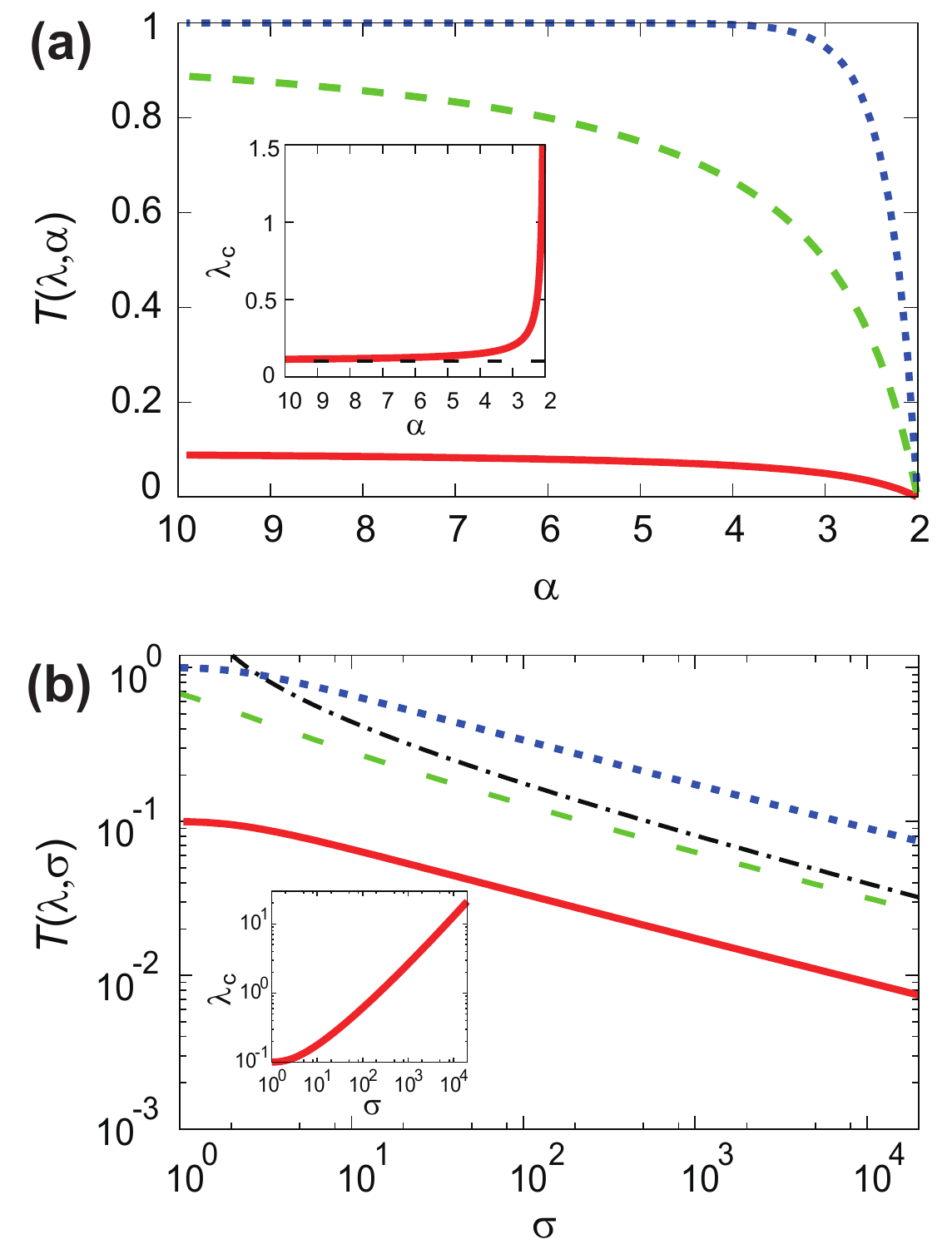}
\caption{
Transmissibility $T$ for (a) the power-law and (b) the lognormal $P(\tau)$ with
$\lambda=10^{-1}$ (solid line), $10^0$ (dashed line), and $10^1$ (dotted line),
plotted as a function of the power-law exponent $\alpha$ and the standard deviations $\sigma$ 
of $P(\tau)$, respectively. $\tm=1$ in (a).
The dash-dotted line in (b) denotes the asymptotic formula, Eq.~(5), for the lognormal $P(\tau)$.
(Insets) Epidemic threshold $\lambda_c$ with $\kappa=10$, vs.\ $\alpha$ (a) and $\sigma$ (b).}
\end{figure}

\section{Transmissibility and epidemic threshold}
Following renewal theory \cite{feller}, 
the transmissibility for contact dynamics following a renewal process with 
independent, identically distributed $P(\tau)$ 
with finite mean waiting time
and fixed recovery time $\lambda$ can be obtained as 
\begin{eqnarray}
T&=&\int_0^\infty g(\Delta) \int_\Delta^\infty \delta(t_R-\lambda) dt_R d\Delta  \\
&=&1-\int_{\lambda}^{\infty}g(\Delta)d\Delta. \nonumber
\end{eqnarray}
Here $g(\Delta)$ is the so-called generation time distribution~\cite{vazquez}, 
the distribution of time intervals between the moment of infection and
the first following contact activity, which in this case 
is given by the residual waiting time distribution, 
$g(\Delta) = \frac{1}{\langle \tau \rangle} \int_\Delta^\infty P(\tau) d\tau$,
where $\langle\tau\rangle$ is the mean waiting time.
The integral with respect to $t_R$ accounts for
the probability that the node does not recover during the interval $\Delta$.
The cases with general recovery time distribution $p(t_R)$ can be treated 
by replacing the delta function with $p(t_R)$ in Eq.~(1).

In Fig.~1, we show the transmissibility $T$ calculated for two heavy-tailed distributions 
that are widely used to model bursty dynamics \cite{temporal,browsing,amaral},
(i) the power-law distribution with exponent $\alpha$ and minimum waiting time ${\tm}$, 
\be P_{PL}(\tau)=\frac{(\alpha-1)}{\tm}\left(\frac{\tau}{\tm}\right)^{-\alpha}\ee
for $\tau>{\tm}$, and $P_{PL}(\tau)=0$ otherwise; and
(ii) the lognormal distribution with unit mean, $\langle \tau\rangle=1$, and variance $\sigma^2$ to focus on effect of the temporal heterogeneity, 
\begin{eqnarray} P_{LN}(\tau)=\frac{1}{\tau\sqrt{2\pi \ln(1+\sigma^2)}}
\exp\left[-\frac{\left(\ln\tau+\ln\sqrt{1+\sigma^2}\right)^2}{2\ln(1+\sigma^2)}\right]. 
\end{eqnarray}
In both cases, the transmissibility $T$ decreases with the contact dynamics' heterogeneity, dictated by either
the power-law exponent $\alpha$ or the variance $\sigma^2$ of $P(\tau)$, respectively.
It even vanishes as $\alpha$ approaches $2$ (Fig.~1a) 
or as the variance diverges (Fig.~1b), respectively. 
This result clearly demonstrates that the heavy-tailed contact dynamics can drastically 
suppress the epidemic spreading.  

For a power-law $P(\tau)$, Eq.~(2), the transmissibility can be calculated explicitly.
It reads
\be\label{q}
T_{PL}(\lambda;\alpha)=
\left\{
\begin{array}{ll}
\frac{(\alpha-2)}{(\alpha-1)}\frac{\lambda}{\tm}, & \lambda\le\tm\\
1-\frac{1}{\alpha-1}\left(\frac{\tm}{\lambda}\right)^{\alpha-2}, & \lambda>\tm.\\
\end{array}
\right.
\ee 
Evidently, $T_{PL}$ decreases as $\alpha$ decreases and 
vanishes as $T_{PL}\sim (\alpha-2)$ as $\alpha\to2$.
For general $P(\tau)$, $T$ may not always be obtained in a simple form. 
Yet its asymptotic behavior can be more accessible for many cases.
For example, for the lognormal waiting time distribution with unit mean, Eq.~(3), 
the residual waiting time distribution is obtained as 
\begin{equation}
g_{LN}(\Delta;\sigma)=\frac{1}{2}\left[1+\textrm{erf}\left(\frac{-\ln\left(\Delta/\sqrt{1+\sigma^2}\right)}{\sqrt{2 \ln(1+\sigma^2})}\right)\right] \nonumber
\end{equation}
, where $\textrm{erf}(x)$ denotes the error function.
Using properties of the error function, one obtains the leading asymptotic behavior of $T$ for large $\sigma$ as
\be T_{LN} \sim \sigma^{-1/4}/\sqrt{\ln(\sigma)} {}, \ee
vanishing algebraically with $\sigma$ (Fig.~1b). 

Once $T$ is obtained, the epidemic threshold can be readily obtained for the process on
uncorrelated tree-like networks by mapping to a branching process \cite{harris}.
From the criticality condition of the branching process, the condition for the epidemic outbreak
is written as $R=T\kappa>1$, where $\kappa$ is the average branching number, given by 
the expected number of neighbors of an infected node excluding the parent node.
For an uncorrelated network $\kappa$ is given by the expected remaining degree of a node 
reached by following a randomly chosen link \cite{newman}, that is
$\kappa=\sum_k (k-1)kP(k)/\langle k\rangle=(\langle k^2\rangle-\langle k\rangle)/\langle k\rangle$. 
For power-law $P_{PL}(\tau)$, Eq.~(2), the epidemic threshold $\lambda_c$ is therefore explicitly obtained as
\be \lambda_{c, PL} = \left\{\begin{array}{ll} 
\frac{\tm (\alpha-1)}{\alpha-2} \frac{1}{\kappa} &  (\lambda_c\le\tm), \\
\tm \left[(\alpha-1)\left(1-\frac{1}{\kappa}\right)\right]^{-1/{(\alpha-2)}} & (\lambda_c>\tm). \end{array}\right.
\ee
Starting from $\lambda_{c}=\tm/\kappa$ in $\alpha\to\infty$ limit 
(denoted by the dashed horizontal line in the inset of Fig.~1a),
$\lambda_c$ increases as $\alpha$ decreases and eventually diverges as $\alpha\to2$ (Fig.~1a, inset).
For $\alpha=2$ the epidemic outbreak cannot take place for any finite $\lambda$, implying that only unrecoverable diseases ($\lambda=\infty$) can spread through the population.
Similarly, $\lambda_c$ diverges with $\sigma$ for the lognormal $P_{LN}(\tau)$ (Fig.~1b, inset).

To obtain $\rho_{\infty}$, one can apply the generating-function method \cite{newman}
based on the mapping to bond percolation in which each bond is randomly occupied with probability $T$.

\begin{figure}
\includegraphics[width=\linewidth]{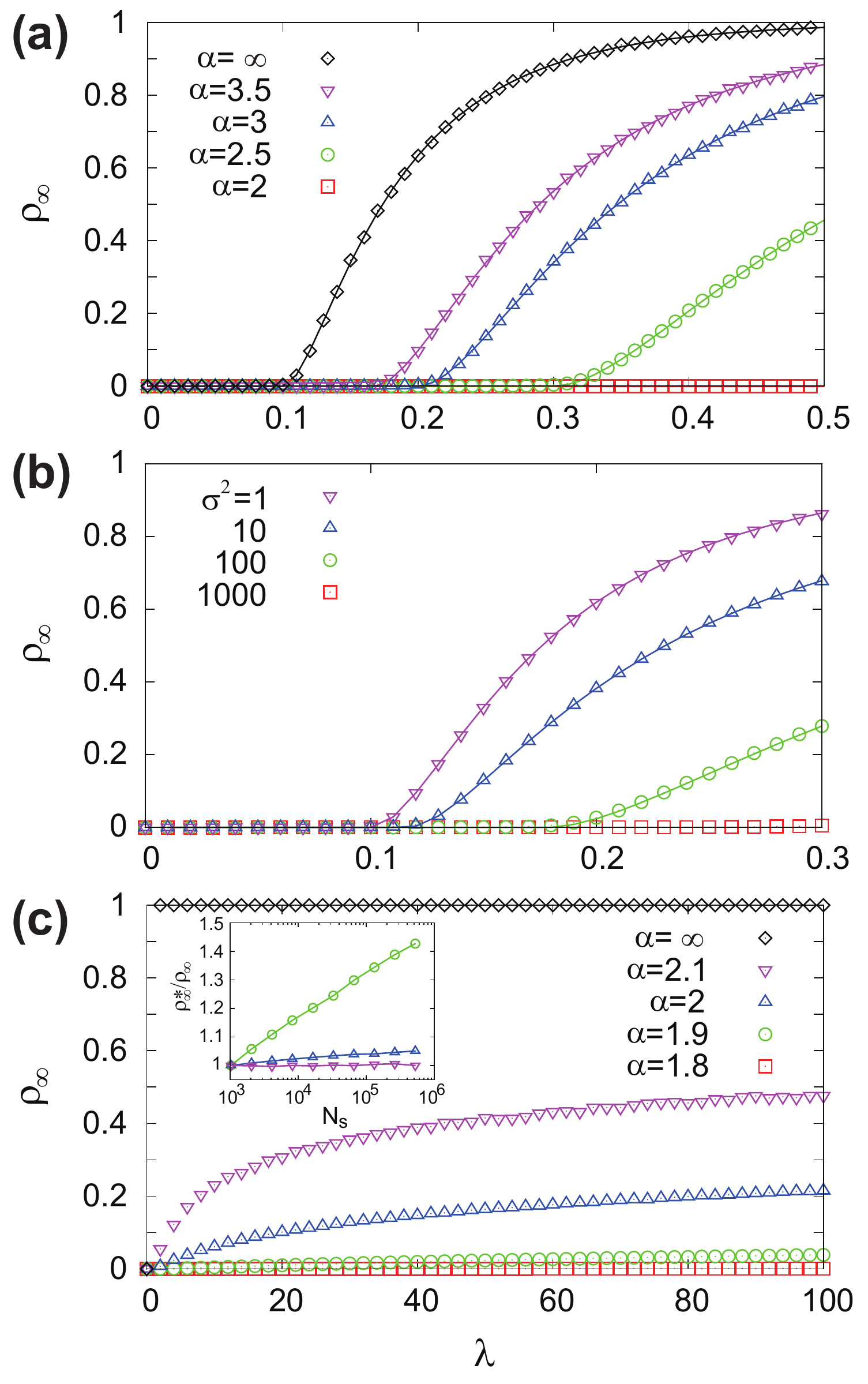}
\caption{
Plots of the final density of recovered nodes (the expected outbreak size) $\rho_{\infty}$ vs.\
the recovery time $\lambda$, of the SIR models.
(a), (b) Results on ER networks with mean degree $\langle k \rangle =10$ and $N=10^4 $,
for (a) the power-law and (b) the lognormal $P(\tau)$.
(c) Results on SF networks with $\gamma=2.5$ and $N=10^4$ with the power-law $P(\tau)$.
Symbols denote numerical simulation results and lines in (a), (b) denotes the theoretical curves,
in excellent agreement with each other.
(Inset) Inverse of the numerically simulated $\rho_{\infty}$ with $\lambda=10$ (rescaled by its value $\rho_{\infty}^*$ for $N_s=10^3$),
plotted against the number of samplings of generation times $N_s$. 
For $\alpha=2.1$ it remains constant, whereas it increases logarithmically
for $\alpha=2.0$ and $1.9$ (same symbols as in main panel). Therefore, as $N_s\to\infty$, $\rho_{\infty}$ is expected to vanish
for $\alpha\le2$, as predicted by the theory.
}
\end{figure}

\section{Numerical simulations} 
We test the validity of the analytical predictions on two random network models,
the Erd\H{o}s-R\'enyi (ER) random graphs \cite{er} and the static model of scale-free (SF) graphs \cite{static}.
The numerical simulation runs as follows.
Given a network of $N$ nodes, initially all nodes in the network 
are susceptible except for one infected node, chosen at random, as a seed node.
Each connected pair of nodes independently makes contacts following a renewal process 
with the  waiting time distribution $P(\tau)$. For the case of power-law $P(\tau)$, Eq.~(2), $\tau$ can be efficiently sampled using the transformation method~\cite{monte}.
The timing of the first contact, however, has to be sampled differently since it is given by the generation time $\Delta$.
We sampled the first contact time as follows. First we constructed a long sequence of contacts following $P(\tau)$, with the number of contacts $N_s$ to be typically ${\cal O}(10^5)$. Then we chose a random time point within the contact sequence and obtained the corresponding generation time.
Whenever an infected agent make a contact with a susceptible neighbor, 
the disease spreads, turning the susceptible node into infected. 
Along the way, each infected node recovers after a fixed recovery period, $\lambda$. 
The process proceeds until there remains no infected agents in the network,
and the final fraction of recovered nodes $S$ is measured.
The ensemble-averaged value of $S$ over independent runs 
gives the expected outbreak size $\rho_{\infty}$.

On ER networks, we show the numerical simulation results with both the power-law 
and lognormal $P(\tau)$, together with the theoretical curves obtained by the aforementioned generating function method (Fig.~2a,b).
The theoretical predictions are in excellent agreement with the numerical simulations.
The epidemic outbreak size consistently decreases and the epidemic threshold diverges  
with the strength of temporal heterogeneity of contact dynamics, 
dictated by $\alpha$ approaching 2 (Fig.~2a) or diverging $\sigma$ (Fig.~2b).

On SF networks with asymptotic power-law degree distribution $P_d(k)\sim k^{-\gamma}$,
it is well-known that the epidemic spreading is facilitated to the extent that 
the epidemic threshold vanishes in the limit of infinite network size when
$\gamma\le3$, as $\kappa$ diverges with $N$ \cite{vespig}.
To verify the impact of temporal heterogeneity in such a case,
we perform extensive numerical simulations with power-law $P(\tau)$ on the SF network with $\gamma=2.5$ (Fig.~2c).
The epidemic outbreak size decreases as $\alpha$ decreases, 
meaning that the temporal heterogeneity can hinder epidemic spreading also in SF networks.
As long as $\alpha>2$, however, $\lambda_c\approx0$, that is, the epidemic outbreak occurs
for any nonzero $\lambda$. In this sense, the network heterogeneity dominates over the
temporal heterogeneity, when $\alpha>2$.
For $\alpha\le2$, however, the temporal heterogeneity can dominate over network heterogeneity to suppress epidemic outbreaks.
In numerical simulation, $\rho_{\infty}$ is obtained to be nonzero, albeit small, for $\alpha=2$ (and even for $\alpha=1.9$),
which is to be attributed as numerical artifact due to finite number of samplings for the time to first contact from $g(\Delta)$
when $\alpha\le2$.
Indeed, $\rho_{\infty}$ is found to decay as the number of samplings is increased,
and thus expected to vanish in the infinite-time limit even for SF networks, as predicted by the theory (Fig.~2c, inset).

\section{Effect of the finite cutoff timescale}
So far, we have assumed that there is no cutoff in the maximum waiting time in $P(\tau)$.
In reality, however, contact dynamics mediating the spreading process 
takes place over a finite time window, bounded, for example, by an individual's lifespan.
We examine the effect of such a cutoff timescale set by the maximum waiting time
on the epidemic outbreak. 
With the cutoff waiting time $\tM$, the generation time distribution is given by 
$g(\Delta)=\frac{1}{\langle \tau \rangle} \int_\Delta^{\tM} P(\tau) d\tau$.

Let us now take $P(\tau)$ to be a power law with exponent $\alpha$ in the range $(\tm, \tM)$.
For $\alpha>2$, $\tM$ plays only a minor effect in the transmissibility,
negligible for large $\tM$.
On the other hand, for $1<\alpha<2$, reportedly corresponding to a number of human activities \cite{browsing},
$\tM$ attains the dominant role in $T(\lambda,\alpha)$, whose leading contribution can be written apart from $\alpha$-dependent proportionality factor as
\be 
T(\lambda;\alpha)\sim\left\{\begin{array}{ll}
\lambda/\tM^{(2-\alpha)} & (\lambda<\tm),\\
(\lambda/\tM)^{2-\alpha} & (\lambda>\tm). 
\end{array}\right.
\ee
Finally, the epidemic threshold $\lambda_c$ depends on $\tM$ as
\be \lambda_c \sim\left\{\begin{array}{ll}
\tM^{2-\alpha}\lambda_{c, \mathrm{P}} & (\lambda<\tm),\\
\tM\lambda_{c, \mathrm{P}}^{1/(2-\alpha)} & (\lambda>\tm). 
\end{array}\right.
\ee
where $\lambda_{c, \mathrm{P}}$ denotes 
the epidemic threshold for Poisson contact dynamics (exponential $P(\tau)$).
The predicted dependence of $\lambda_c$ on $\tM$ is well supported by the numerical simulations (Fig.~3). 
This result shows that the more heavy-tailed (smaller $\alpha$) the contact dynamics is, the larger is the impact of long but finite waiting times. 

\begin{figure}
\includegraphics[width=\linewidth]{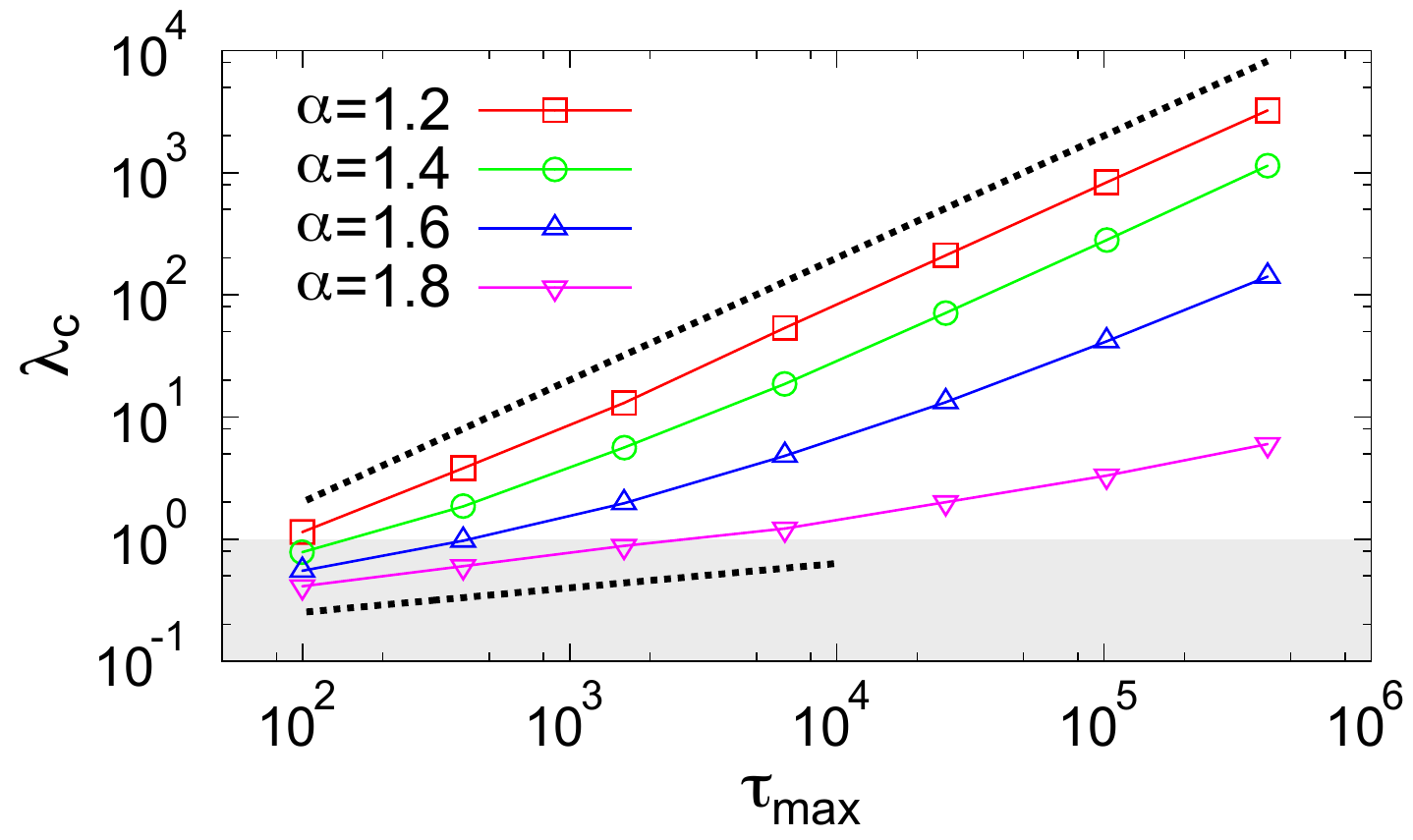}
\caption{
The epidemic threshold $\lambda_c$ vs.\ the maximum waiting time $\tM$
in the power-law waiting time distribution with various exponent $\alpha$, 
obtained from numerical simulations of the SIR model 
on the scale-free network with $\gamma=2.5$ and $N=10^4$.
Dotted lines have slope $1.0$ (top) and $0.2$ (bottom),
shown for comparison with the theoretical prediction, Eq.~(8).
Two regimes are separated by the minimum $\tau$, $\tm=1$, indicated by the shade area.
}
\end{figure}

\section{Summary} 
To summarize, we have shown both analytically and numerically 
that epidemic outbreaks can be strongly suppressed by the heavy-tailed contact dynamics.
Applying renewal theory, we have derived the transmissibility $T$ and epidemic threshold $\lambda_c$ 
for contact dynamics following power-law and lognormal waiting time distributions. 
It is shown explicitly that $T$ vanishes (consequently, $\lambda_c$ diverges) as $\alpha\to2$
or $\sigma\to\infty$, respectively, 
which are specific instances of the general condition for diverging $\lambda_c$, 
given by $\int_{\lambda}^{\infty}g(\Delta)d\Delta=1$ for any finite $\lambda$. 
As such, temporal heterogeneity is found to exert opposite effect to network heterogeneity, 
and so they compete with each other.
Finally, it is noteworthy that although we have specifically formulated our analysis 
with fixed recovery time, the main result of 
suppressing effect of temporal heterogeneity would apply under more general epidemic scenarios.
For example, we have observed qualitatively similar suppression of epidemic dynamics with exponentially distributed recovery times \cite{miller,kenah} and the susceptible-infected-susceptible-type dynamics \cite{anderson}.

\acknowledgements
We thank S.-C.~Park for useful discussions. 
This work was supported by Basic Science Research Program through NRF grant 
funded by the MEST of Korea (No.\ 2011-0014191).

\end{document}